\documentclass[conference,10pt]{IEEEtran}

\usepackage[T1]{fontenc}
\usepackage[utf8]{inputenc}
\usepackage{microtype}
\usepackage{graphicx}
\usepackage{booktabs}
\usepackage{array}
\usepackage{tabularx}
\usepackage{listings}
\usepackage{fancyvrb}
\usepackage{xcolor}
\usepackage{tikz}
\usetikzlibrary{positioning, arrows.meta, shapes.geometric, fit, calc,
                backgrounds, decorations.pathreplacing}
\usepackage{pgfplots}
\pgfplotsset{compat=1.18}
\usepackage{amsmath}
\usepackage{url}
\usepackage[hidelinks]{hyperref}
\usepackage{cleveref}

\definecolor{lstbg}{gray}{0.97}
\definecolor{lstkw}{rgb}{0.10,0.30,0.65}
\definecolor{lstcm}{rgb}{0.30,0.55,0.30}
\definecolor{lststr}{rgb}{0.65,0.10,0.10}

\lstdefinestyle{cstyle}{
  language=C,
  basicstyle=\ttfamily\footnotesize,
  keywordstyle=\color{lstkw}\bfseries,
  commentstyle=\color{lstcm}\itshape,
  stringstyle=\color{lststr},
  backgroundcolor=\color{lstbg},
  frame=single,
  rulecolor=\color{lstbg},
  breaklines=true,
  showstringspaces=false,
  columns=fullflexible,
  numbers=none,
  xleftmargin=2pt,
  xrightmargin=2pt,
  aboveskip=4pt,
  belowskip=4pt,
}
\lstdefinestyle{shellstyle}{
  basicstyle=\ttfamily\scriptsize,
  backgroundcolor=\color{lstbg},
  frame=single,
  rulecolor=\color{lstbg},
  breaklines=true,
  columns=fullflexible,
  numbers=none,
  xleftmargin=2pt,
  xrightmargin=2pt,
  aboveskip=4pt,
  belowskip=4pt,
}
\title{SynapticOS: An Inference-First Runtime Architecture for
Neural Processing Units on Resource-Constrained Microcontrollers}

\author{\IEEEauthorblockN{Dimitrios Kafetzis}
\IEEEauthorblockA{SynapticOS Project, Hamburg, Germany}}

\begin{document}
\maketitle

\begin{abstract}
Microcontrollers with on-die neural processing units (NPUs) have
become mainstream, but the system software hosting them has not: the
production combinations of Zephyr or FreeRTOS with TensorFlow Lite
Micro treat AI inference as an application-layer library, leaving
memory fragmentation, accelerator-state hygiene, and model-lifecycle
guards as recurring application-developer concerns. We present the
Phase~1 foundation of SynapticOS, an open-source runtime built on
Zephyr that treats inference as a first-class workload. The foundation
contributes four cooperating subsystems: (1)~a tensor-aware bump
allocator with 16-byte DMA-aligned persistent and ephemeral lifetimes
sharing a single arena, achieving constant-time allocation
($\sim$154 cycles per call, $\sim$78{,}000 allocations per second at
150~MHz, invariant across tensor sizes) with zero fragmentation by
construction; (2)~a four-state hardware abstraction layer for the NPU
and DSP, implemented by both a deterministic software stub (for
continuous integration under QEMU) and a Neutron-flavoured backend
(for the NXP MCXN947); (3)~a three-state model lifecycle registry
with duplicate-name detection, idempotent load/unload, and hot-swap
guards; and (4)~a four-mark cycle-accurate profiler. We evaluate on
the NXP FRDM-MCXN947 (dual Cortex-M33 at 150~MHz) and on the
\texttt{qemu\_cortex\_m3} emulator. Build footprints are 67~KB flash
/ 184~KB SRAM on FRDM (with shell, 128~KB arena) and 24~KB flash /
28~KB SRAM on QEMU (no shell, 8~KB arena). End-to-end inference
brackets through the deterministic stub kernel measure
1{,}038~\textmu s on FRDM and 781~\textmu s on QEMU for a
$16\!\times\!16\!\times\!3$ INT8 input; these are baseline overhead
numbers, not measurements of the Neutron silicon, which is exercised
by the real SDK invoke path scheduled for Phase~2. A 61-test suite
spanning 10 ZTEST suites passes 100\% in 6.6~s on the CI emulator
path. SynapticOS is released under Apache~2.0 at
\url{https://github.com/Dimitrios-Kafetzis/SynapticOS}.

\end{abstract}


\section{Introduction}
\label{sec:intro}

A neural processing unit on a microcontroller is no longer exotic.
NXP's MCXN9 family integrates the eIQ Neutron NPU on a dual-Cortex-M33
package retailing at roughly fifteen US dollars on a reference
board~\cite{nxp-mcxn947, nxp-neutron}; Arm's Ethos-U55 ships as a
licensable accelerator paired with Cortex-M55 and M85
cores~\cite{ethos-u}; STMicroelectronics, Renesas, and Espressif have
each announced or shipped comparable on-die accelerators in their
recent silicon. Within roughly two product cycles, INT8 inference at
single-digit-watt power budgets has migrated from application
processors down into the same class of devices that runs a doorbell
or a thermostat.

The system software has not made the corresponding move. The two
production RTOS platforms with the largest install bases ---
Zephyr~\cite{zephyr} and FreeRTOS~\cite{freertos} --- both treat AI
inference as an application-layer library: a developer links against
TensorFlow Lite for Microcontrollers~\cite{tflm} or
CMSIS-NN~\cite{cmsis-nn}, allocates a tensor arena out of the
standard heap, and invokes the accelerator through a thin vendor SDK
wrapper. This works, but it leaves three classes of problem to the
application developer to solve, again, on every project:

\begin{itemize}
\item \textbf{Memory.} A general-purpose heap interleaves model
weights (which live for the lifetime of a loaded model) and
activations (which live for one forward pass), producing the
classical checkerboard fragmentation pattern that limits how many
inferences a deployed device can run before requiring a reboot.
\item \textbf{Hardware abstraction.} The NPU on an MCXN947 board, the
Ethos-U on a Cortex-M85 board, and the software fallback on an
ESP32-S3 all need to look the same to the inference pipeline, and
they do not: each ships with its own SDK conventions, its own
async-completion contract, and its own power-management story.
\item \textbf{Lifecycle.} A model has a state (registered, loaded,
swapped, retired) that needs to be tracked separately from the
accelerator state, with guards against the obvious misuses (load
while busy, swap while loaded, unregister while loaded). No
mainstream MCU runtime ships this surface today.
\end{itemize}

The micro-TVM compiler stack~\cite{microtvm} takes a different
approach: schedule the entire inference graph ahead of time and emit
straight-line code with statically planned tensors. This is the right
choice when the model is fixed and known at build time, but it is the
wrong abstraction for a device that needs to load, update, or hot-swap
models in the field, and it offers no help with the hardware-state and
lifecycle problems above. CMSIS-NN occupies a similar niche one layer
down: efficient kernels with caller-managed memory and no runtime
notion of a model.

\subsection{Inference-First Design}
\label{sec:intro:firstprinciples}

The starting position of this work is that the three problems above
are not application-layer concerns. They are workload-class concerns:
they recur on every project because the workload --- repeated forward
passes through a quantised neural network --- has memory, hardware,
and lifecycle patterns that differ structurally from the control loops
and sensor-polling tasks that the existing RTOS abstractions were
shaped around. An OS that treats inference as a first-class workload
can name those patterns directly and solve them once.

The three observations we build on are:

\begin{enumerate}
\item \textbf{Tensor memory has two lifetimes, not one.} Weights are
loaded once per model and read on every forward pass; activations are
written once per forward pass and discarded. Allocating both from a
single free-list forces a policy choice that is wrong for at least
one of them; splitting the two and using a bump allocator on each
makes both choices trivially correct.
\item \textbf{Accelerator state is a state machine, not a flag.}
\texttt{IDLE}, \texttt{BUSY}, \texttt{SUSPENDED}, and \texttt{ERROR}
are not implementation details of a particular SDK --- they are the
shared vocabulary that the model registry, the scheduler, and the
power manager all need to consult. Putting that state in a single
HAL means every layer above the HAL gets a coherent view by
construction.
\item \textbf{A deterministic software fallback unlocks
continuous integration.} Most RTOS-plus-NPU projects ship without
unit tests for the NPU path because the test bench has no NPU. A
stub backend with bit-identical output for identical input solves
this without giving up on the real-hardware path.
\end{enumerate}

These observations are the design axes for SynapticOS Phase~1. The
work reported here is the foundation layer --- memory, HAL, registry,
profiling, shell, and test infrastructure --- on which the Phase~2
inference engine (TFLM integration, layer-granularity preemption,
real Neutron invoke) and the later phases (dual-core IPC, OTA, fault
recovery) are built.

\subsection{Contributions}
\label{sec:intro:contributions}

The Phase~1 contributions are:

\begin{enumerate}
\item A \textbf{tensor-aware memory architecture}
(\cref{sec:memory}): a bump allocator with persistent and ephemeral
regions sharing a single tensor arena and a separate scratch pool,
with 16-byte DMA alignment by construction and constant-time
allocation independent of tensor size, validated at $\sim$154
cycles per allocation and $\sim$78{,}000 allocations per second on
\texttt{qemu\_cortex\_m3} (\cref{sec:eval:mem}).
\item A \textbf{state-machine NPU HAL with deterministic software
fallback} (\cref{sec:hal}): a four-state interface implemented by
both a software-stub backend (linked for QEMU) and a Neutron-flavoured
backend (linked for FRDM-MCXN947), selected at link time with no
runtime dispatch overhead.
\item A \textbf{model lifecycle registry} (\cref{sec:registry:lifecycle}):
a three-state machine over a fixed-size slot table with duplicate-name
detection, idempotent load/unload, and auto-unload-on-unregister
guards.
\item A \textbf{cycle-accurate profiling surface}
(\cref{sec:registry:profiling}): four mark-points along the
preprocess--invoke--postprocess pipeline backed by
\texttt{k\_cycle\_get\_32()}, with arena-peak capture and an NPU
utilisation ratio. The mark API is in place; wiring it into the
inference path is Phase~2 work and is reported as a known gap.
\item An \textbf{open-source dual-target implementation} of all of
the above: roughly 4{,}100 lines of C across the runtime, frozen
public headers, samples, and tests; 61 ZTEST cases across 10
suites with a 100\% pass rate on the CI emulator path; and a
working FRDM-MCXN947 deployment with a Zephyr shell front-end.
Source is released under Apache 2.0 at
\url{https://github.com/Dimitrios-Kafetzis/SynapticOS}.
\end{enumerate}

\subsection{Scope}
\label{sec:intro:scope}

This paper reports the Phase~1 \emph{foundation} of SynapticOS. The
foundation is deliberately under the inference engine, not the
inference engine itself: the latency numbers in \cref{sec:eval:latency}
are bracketed end-to-end times around a deterministic stub kernel,
not measurements of the Neutron NPU on a real workload. The real
inference engine --- TFLite Micro integration, zero-copy tensor
pipelines, layer-granularity preemption, the real Neutron invoke
path, and per-stage profiler wire-up --- is Phase~2 work. We are
explicit about this throughout (in \cref{sec:eval:setup,sec:eval:latency,sec:registry:profiling,sec:hal:backends})
so that a reader of the latency figures cannot mistake them for
silicon performance.

\subsection{Paper Organisation}
\label{sec:intro:org}

\Cref{sec:related} situates SynapticOS against the existing landscape
of RTOS-plus-AI integrations. \Cref{sec:memory,sec:hal,sec:registry}
present the three core subsystems (memory, HAL, registry plus
profiling). \Cref{sec:eval} evaluates build footprint, latency,
allocator performance, and test coverage on both target platforms.
\Cref{sec:discussion} discusses limitations and the roadmap;
\cref{sec:conclusion} concludes.

\section{Related Work}
\label{sec:related}

SynapticOS sits at the intersection of three lines of existing work: the
real-time operating systems that host embedded AI today, the tensor
memory managers that those AI frameworks ship internally, and the
hardware abstractions that the larger AI runtimes use to talk to
accelerators. We do not claim novelty against any single line in
isolation; what is missing is the OS-level synthesis of all three for
the MCU-plus-NPU class of device, and it is that gap we address.

\subsection{RTOS Hosts for Embedded AI}
\label{sec:related:rtos}

\Cref{tab:rtos-compare} summarises how the major embedded RTOS and
runtime stacks expose AI to the application. In every existing
production combination the AI library is a peer of (or a client of)
the RTOS, not an integrated subsystem: the runtime supplies a heap and
a scheduler, the library supplies an inference pipeline and an arena
on top of that heap, and the application stitches them together.

\begin{table*}[t]
\caption{Embedded AI integration patterns.}
\label{tab:rtos-compare}
\centering
\renewcommand{\arraystretch}{1.2}
\begin{tabular}{@{}lllll@{}}
\toprule
\textbf{System} & \textbf{AI support} & \textbf{Memory model} &
\textbf{Accelerator HAL} & \textbf{Model lifecycle} \\
\midrule
Zephyr + TFLM~\cite{zephyr,tflm} & app-layer library & TFLM arena over
heap & vendor SDK (per-board) & app-managed \\
FreeRTOS + TFLM~\cite{freertos,tflm} & app-layer library & TFLM arena
over heap & vendor SDK (per-board) & app-managed \\
$\mu$TVM~\cite{microtvm} & compiler-generated runtime & statically
planned & AOT-bound to one target & one model per build \\
ARM CMSIS-NN~\cite{cmsis-nn} & kernel library & caller-supplied
buffers & none & none \\
\midrule
\textbf{SynapticOS} (this work) & \textbf{native runtime} &
\textbf{tensor-aware arena} & \textbf{state-machine HAL} &
\textbf{registry + guards} \\
\bottomrule
\end{tabular}
\end{table*}

The Zephyr-plus-TFLM and FreeRTOS-plus-TFLM combinations are the
de-facto production stack for shipping MCU AI today. They work and
they scale: TFLM has been deployed on hundreds of millions of devices
according to the project's own figures, and the broader MLPerf~Tiny
benchmark suite~\cite{banbury21mlperftiny} centres on this class of
deployment. The pattern's limitation is
not technical inadequacy --- it is that every project re-derives the
same arena-sizing decisions, the same accelerator state hygiene, and
the same model-lifecycle guards on top of a runtime that does not name
those concerns. Two consequences of this are visible in the field:
TFLM's static arena requires the developer to over-allocate against
the largest expected activation tensor (because the framework has no
ephemeral / persistent distinction to exploit), and accelerator
suspension and resume on power-managed devices is the application's
problem because the framework has no concept of accelerator state.
SynapticOS keeps TFLM in scope as a Phase~2 integration target ---
the goal is not to displace it but to give it (or any other inference
library) an OS layer that already solves the recurring problems.

The $\mu$TVM stack~\cite{microtvm} takes the orthogonal route of
ahead-of-time compilation: an entire inference graph is lowered to
straight-line C with statically planned tensor offsets. For a fixed,
single-model build this is excellent and competitive with hand-tuned
kernels, but the abstraction breaks if the device needs to load,
hot-swap, or OTA-update models in the field, and the compiler offers
no help with hardware-state management or with the shell-level
introspection that real-world deployments need for debugging.

CMSIS-NN~\cite{cmsis-nn} is one layer below the others: a library of
optimised neural-network kernels (convolutions, fully-connected,
pooling, activation) that callers wire together by hand. It is the
right comparator for our DSP HAL (\cref{sec:hal:dsp}) but explicitly
not a runtime; it has no notion of memory ownership, no model lifecycle,
and no accelerator HAL. Our DSP HAL adopts CMSIS-NN-style primitive
contracts (caller-supplied buffers, separate scale and zero-point for
quantised inputs) without inheriting its lack of OS-level integration.

Nuttx, Mbed-OS, and the various vendor-specific microcontroller
runtimes (e.g.\ MCUXpresso, ESP-IDF) follow the same library-attached
pattern as Zephyr-plus-TFLM and FreeRTOS-plus-TFLM and are omitted
from \cref{tab:rtos-compare} for brevity. None of them ship an
OS-native tensor-aware allocator or an OS-native NPU state machine
either.

\subsection{Tensor Memory Management}
\label{sec:related:mem}

The TFLite Micro arena allocator~\cite{tflm} is the closest existing
work to SynapticOS's memory subsystem. It uses a bump-pointer scheme
within a caller-supplied byte buffer, with two regions: persistent (head
of arena, growing up) and non-persistent / temporary (tail of arena,
growing down). On the surface this resembles \cref{sec:memory:layout},
but two differences matter for an OS-level use:

\begin{itemize}
\item TFLM's arena is owned by the inference interpreter, sized at
interpreter construction, and shared across no other consumers. An
OS-level allocator has to serve callers that are not inside the
interpreter (e.g.\ a sensor pre-processor, an IPC marshaller), which
is why SynapticOS exposes the arena through a small public API
(\cref{lst:memapi}) rather than as a hidden interpreter member.
\item TFLM has no separate scratch pool. Per-operator scratch is
carved out of the same non-persistent region, which forces the
operator code to reason about lifetime against neighbouring tensors.
A separate, independently reset scratch pool (\cref{sec:memory:layout})
keeps that reasoning local to the operator.
\end{itemize}

Pool-of-pools and slab allocators in production RTOS kernels
(Zephyr's \texttt{k\_mem\_slab}, FreeRTOS's heap\_4) solve a different
problem: they trade fragmentation for fixed-size restriction. Tensor
allocations are not fixed-size, so a slab allocator either wastes
significant memory or requires a slab class per expected tensor size,
which loses the determinism advantage it was chosen for.

The bump-pointer / linear / region allocator family itself is folklore
that long predates AI workloads; it is the standard idiom in
high-throughput game engines, compilers, and short-lived-arena server
patterns~\cite{gameengine, regionalloc}. SynapticOS's contribution is
not the bump-pointer choice itself --- it is the lifetime
classification and the OS-level API that surfaces it.

\subsection{Hardware Abstraction for Accelerators}
\label{sec:related:hal}

On the application-processor and mobile side, the field has converged
on the Execution Provider (EP) and device-API patterns. ONNX
Runtime~\cite{onnxrt} dispatches operators to EPs at graph-execution
time; TVM's device API~\cite{tvm} exposes a thin C interface for
allocator, stream, and kernel launch operations that is implemented
per-backend; Android NNAPI~\cite{nnapi} is conceptually similar at the
mobile-application layer. All three are too heavy for a Cortex-M class
device: ONNX Runtime's footprint alone is in the multi-megabyte range
even after aggressive subsetting, and the dispatch path assumes a
device that can afford virtual-function overhead per operator.

At the MCU layer the prevailing approach is the opposite extreme:
each vendor's SDK ships its own NPU driver with its own state model,
its own buffer-ownership contract, and its own async-completion API.
The NXP eIQ Neutron SDK, the Arm Vela compiler and Ethos-U driver,
the Renesas DRP-AI driver, and the ST X-CUBE-AI runtime are all
internally consistent and all mutually incompatible. The
\texttt{syn\_hal\_npu.h} interface (\cref{sec:hal}) is deliberately
narrow --- a state enum, capability struct, lifecycle functions, and a
single synchronous-plus-async invoke pair --- precisely so that it can
sit above all of these SDKs without paying ONNX-Runtime-class overhead.
The dual-backend implementation we ship for Phase~1 (stub and Neutron)
demonstrates that the interface is implementable; the Phase~2 work
adds the real Neutron-SDK invoke path behind the same surface and
the Phase~3 work adds the asymmetric-multiprocessing dimension.

\section{Tensor-Aware Memory Architecture}
\label{sec:memory}

A neural-network inference kernel has a memory-access pattern that is
fundamentally different from the control-loop and sensor-polling workloads
that traditional RTOS allocators are tuned for. Weights are loaded once and
read many times; activations are written and overwritten on every forward
pass; intermediate scratch is needed by individual operators (FFT twiddle
factors, softmax denominators) and discarded immediately. SynapticOS exposes
these three lifetimes directly to the allocator instead of forcing them
through a single general-purpose heap.

\subsection{Design Rationale}
\label{sec:memory:rationale}

A general-purpose dynamic allocator --- whether \texttt{newlib malloc},
\texttt{k\_malloc}, or a free-list heap --- has three properties that make it
a poor fit for inference on a microcontroller:

\begin{itemize}
\item \textbf{Fragmentation under heterogeneous lifetimes.} Weights live for
the lifetime of a loaded model; activations live for one forward pass.
Allocating both from a single free-list pool interleaves long-lived and
short-lived blocks, producing the checkerboard-fragmentation pattern
that Wilson et al.~\cite{wilson95mem} document as defeating any
first-fit or best-fit policy under a fixed memory budget.
\item \textbf{Unpredictable latency.} Free-list traversal is bounded only by
the current pool topology, which is workload-dependent. A real-time
inference deadline cannot tolerate a worst-case allocation that is two or
three orders of magnitude slower than the typical case.
\item \textbf{No alignment guarantee.} NPU and DMA engines on Cortex-M class
hardware typically require 16-byte (sometimes 32-byte) alignment for
input and output tensors. Standard heap implementations only guarantee
4- or 8-byte alignment and require callers to over-allocate and align
manually, which both wastes memory and complicates the surrounding code.
\end{itemize}

The Phase~1 allocator addresses all three by construction: lifetime
classification eliminates fragmentation, the bump-pointer algorithm
guarantees $O(1)$ allocation, and every tensor descriptor and payload is
placed on a 16-byte boundary by an unconditional \texttt{ALIGN\_UP} at the
allocation site (\texttt{syn\_mem.c:23,112}).

\subsection{Arena Layout}
\label{sec:memory:layout}

\begin{figure}[t]
\centering
\resizebox{\columnwidth}{!}{%
\begin{tikzpicture}[
  font=\small,
  region/.style={
    rectangle,
    draw=black!70,
    thick,
    minimum height=18mm,
    text width=#1,
    align=center,
    inner sep=2pt,
    outer sep=0pt,
  },
  arrow/.style={-Latex, semithick, black!75},
  tick/.style={black!55, thin},
  brace/.style={decorate, decoration={brace, amplitude=4pt, raise=2pt},
                semithick, black!65},
  sub/.style={font=\scriptsize\itshape, text=black!60},
]

\node[region=14mm, fill=blue!12]  (persi)   at (0,0)
  {persistent};
\node[region=18mm, fill=orange!18](eph)     [right=0mm of persi]
  {ephemeral};
\node[region=32mm, fill=black!4]  (free)    [right=0mm of eph]
  {free};
\node[region=20mm, fill=green!15] (scratch) [right=0mm of free]
  {scratch\\pool\\[1pt]
   \textit{\textcolor{black!60}{\footnotesize (resets with}}\\
   \textit{\textcolor{black!60}{\footnotesize ephemeral)}}};

\draw[arrow] ($(persi.south west) + ( 2mm, 4mm)$)
           -- ($(persi.south east) + (-2mm, 4mm)$);
\draw[arrow] ($(eph.south west)   + ( 2mm, 4mm)$)
           -- ($(eph.south east)   + (-2mm, 4mm)$);

\draw[brace]
  (persi.north west) -- (free.north east)
  node[midway, above=6pt, font=\footnotesize]
    {tensor region (bump-allocated)};

\draw[tick] (persi.south west)   -- ++(0,-3mm) coordinate (t_base);
\draw[tick] (scratch.south west) -- ++(0,-3mm) coordinate (t_usable);
\draw[tick] (scratch.south east) -- ++(0,-3mm) coordinate (t_total);

\node[below=0pt of t_base,   font=\footnotesize, anchor=north]
  {\texttt{base}};
\node[below=0pt of t_usable, font=\footnotesize, anchor=north,
      align=center]
  {\texttt{base}\\\texttt{+ usable}};
\node[below=0pt of t_total,  font=\footnotesize, anchor=north,
      align=center]
  {\texttt{base}\\\texttt{+ total}};

\end{tikzpicture}%
}
\caption{Tensor arena layout. The arena is a single contiguous block
split into a bump-allocated tensor region and a separate scratch pool.
Within the tensor region, persistent allocations grow up from
\texttt{base}, and ephemeral allocations continue upward where
persistent ends; the scratch pool is reclaimed in lockstep with
\texttt{syn\_mem\_reset\_ephemeral()}. The allocator returns
\texttt{-ENOMEM} when \texttt{persistent\_used~+~ephemeral\_used}
would exceed \texttt{usable~=~total~-~scratch\_size}.}
\label{fig:arena}
\end{figure}
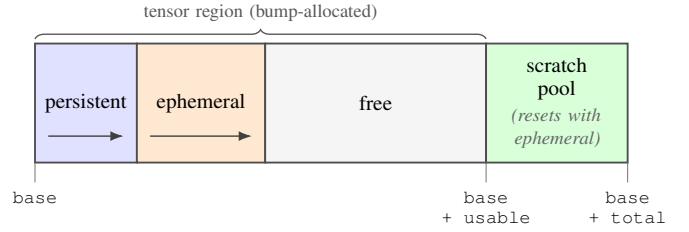

The arena is a single contiguous block of SRAM, partitioned at
initialization into a tensor region and a scratch pool. Within the tensor
region, two bump pointers grow in the same direction starting at the base:
the persistent pointer occupies the low addresses, and the ephemeral
pointer takes over once the persistent allocations have stabilized. The
scratch pool occupies the high end of the arena and is reclaimed at the
same lifetime boundary as the ephemeral region. \Cref{fig:arena} shows
the layout; this is the structure actually implemented in
\texttt{syn\_mem.c}.

The split is parameterized by the Kconfig knob
\texttt{CONFIG\_SYNAPTIC\_SCRATCH\_POOL\_SIZE} (range 1{,}024--65{,}536
bytes); on FRDM the default 128~KB arena is split into a 112~KB tensor
region and a 16~KB scratch pool, and on QEMU an 8~KB arena is split
7~KB / 1~KB. The runtime exposes both occupancies through
\texttt{syn\_mem\_get\_stats()} and the \texttt{syn mem stats} shell
command (\cref{sec:eval:shell}).

\subsection{Allocation Algorithm}
\label{sec:memory:algo}

A tensor allocation proceeds in four steps:

\begin{enumerate}
\item Compute the element count by multiplying the (up to four) shape
dimensions, then multiply by the dtype size (1, 2, or 4 bytes for the
supported INT8/UINT8, INT16/FLOAT16, and FLOAT32 dtypes).
\item Reserve an inline descriptor of size $\lceil \mathrm{sizeof}(\mathtt{syn\_tensor\_t}) \rceil_{16}$
immediately followed by the payload, so the entire allocation is a
single contiguous block. The payload pointer in the returned descriptor
is set to the byte after the padded descriptor.
\item Select the bump pointer for the requested lifetime:
\texttt{persistent\_used} for \texttt{SYN\_MEM\_PERSISTENT} and
\texttt{persistent\_used + ephemeral\_used} for everything else (see
note below on \texttt{SYN\_MEM\_SHARED}).
\item Advance the chosen pointer past the aligned end. If the new end
would exceed the tensor region (\texttt{persistent\_used + ephemeral\_used > usable}),
the call returns \texttt{NULL} and no state mutates; otherwise the
allocation count is incremented, the high-water mark is updated, and
the descriptor is returned.
\end{enumerate}

A call to \texttt{syn\_mem\_reset\_ephemeral()} zeroes both
\texttt{ephemeral\_used} and \texttt{scratch\_used} (so scratch is on the
ephemeral lifetime by construction, not on a separate lifetime) and
increments the reset counter. Persistent state is never touched by a
reset. \texttt{syn\_mem\_tensor\_free()} is a deliberate no-op: the bump
allocator does not track individual frees, and pretending otherwise would
mislead callers.

\paragraph*{A note on \texttt{SYN\_MEM\_SHARED}.} The public header
declares three lifetimes (\texttt{PERSISTENT}, \texttt{EPHEMERAL},
\texttt{SHARED}), but the Phase~1 implementation only distinguishes two:
the lifetime branch in \texttt{arena\_alloc()} checks for
\texttt{PERSISTENT}, and everything else --- including
\texttt{SHARED} --- bumps from the ephemeral pointer. The
\texttt{SHARED} value is reserved for the dual-core IPC region scheduled
for the asymmetric-multiprocessing work in Phase~3; passing it today
yields ephemeral semantics. This is called out here so that callers do
not rely on \texttt{SHARED} surviving an ephemeral reset.

\subsection{API}
\label{sec:memory:api}

The public surface is intentionally small. \Cref{lst:memapi} reproduces
the frozen signatures from \texttt{include/synaptic/syn\_mem.h}; the
scratch pool has a symmetric acquire/release pair where
\texttt{release} is a no-op (release is implicit at the next ephemeral
reset).

\begin{lstlisting}[style=cstyle,caption={Phase~1 memory API
(\texttt{include/synaptic/syn\_mem.h}, frozen).},label={lst:memapi},
captionpos=b]
int  syn_mem_init(void *arena_base, size_t arena_size);
void syn_mem_reset_ephemeral(void);

syn_tensor_t *syn_mem_tensor_alloc(const uint32_t *shape,
                                   uint8_t ndim,
                                   syn_npu_dtype_t dtype,
                                   syn_mem_lifetime_t lifetime);
void syn_mem_tensor_free(syn_tensor_t *tensor);

void *syn_mem_scratch_acquire(size_t size);
void  syn_mem_scratch_release(void *ptr);

int  syn_mem_get_stats(syn_mem_stats_t *stats);
\end{lstlisting}

Quantitative behaviour --- per-allocation cost, size invariance,
fragmentation accounting, and region isolation --- is reported in
\cref{sec:eval:mem}, where the design choices above are validated against
the benchmark suites and the runtime statistics from a real inference on
the FRDM-MCXN947.

\section{Hardware Abstraction Layer}
\label{sec:hal}

The accelerators on a microcontroller class device --- an NPU for
neural-network invoke, a DSP block for vector and tensor primitives,
DMA engines for zero-copy transfers --- have vendor-specific register
maps, lifecycle quirks, and clock and power-domain requirements. A
clean hardware abstraction layer matters for two reasons that go beyond
ordinary portability. First, the upstream Zephyr build must be able to
finish in continuous integration without the silicon being present;
Phase~1 achieves this by linking against software fallback backends on
\texttt{qemu\_cortex\_m3} that expose the same public interface as the
real drivers. Second, the model lifecycle in \cref{sec:registry}
encodes safety guards (e.g.\ ``cannot load while another invoke is in
flight'') that need a single source of truth for accelerator state;
SynapticOS centralises that state in the HAL rather than scattering it
across drivers.

Two HALs are part of the Phase~1 surface: the NPU HAL
(\texttt{syn\_hal\_npu.h}) and the DSP HAL
(\texttt{syn\_hal\_dsp.h}). A DMA HAL is declared but not exercised by
the Phase~1 sample and is therefore omitted from this discussion.

\subsection{NPU State Machine}
\label{sec:hal:state}

The NPU HAL exposes a four-state machine
(\cref{fig:npustate}): \texttt{IDLE}, \texttt{BUSY},
\texttt{SUSPENDED}, and \texttt{ERROR}.\footnote{The implementation
additionally maintains a private \texttt{initialized} boolean. We do
not promote this to a public state because pre-init and post-deinit
look identical to the caller, who only sees \texttt{-EPERM} either way.}
Transitions are precondition-checked at every call site: e.g.\ a
\texttt{load\_model} or \texttt{set\_input} call while the NPU is
\texttt{BUSY} returns \texttt{-EBUSY} without mutating any state, and
\texttt{resume} from any state other than \texttt{SUSPENDED} returns
\texttt{-EINVAL}. These guards prevent the model registry
(\cref{sec:registry}) from being able to corrupt the HAL through a
misordered API call --- a property that becomes especially important
once preemption is added in Phase~2.

\begin{figure}[t]
\centering
\resizebox{\columnwidth}{!}{%
\begin{tikzpicture}[
  font=\scriptsize,
  state/.style={
    rectangle,
    rounded corners=3pt,
    draw=black!70,
    thick,
    minimum width=14mm,
    minimum height=7mm,
    align=center,
    inner sep=2pt,
  },
  entry/.style={state, fill=blue!10},
  active/.style={state, fill=orange!15},
  suspended/.style={state, fill=green!10},
  err/.style={state, fill=red!10, draw=red!50!black, dashed},
  trans/.style={-Latex, thick, black!70},
  errlbl/.style={font=\tiny\itshape, red!50!black},
  lbl/.style={font=\tiny},
]

\node[entry]     (idle)                            {\texttt{IDLE}};
\node[active]    (busy)      [right=12mm of idle]  {\texttt{BUSY}};
\node[suspended] (suspended) [below=8mm of idle]   {\texttt{SUSPENDED}};
\node[err]       (error)     [below=8mm of busy]   {\texttt{ERROR}};

\coordinate (initstart) at ($(idle.west) + (-12mm, 0)$);
\draw[trans] (initstart) -- (idle)
  node[midway, above, lbl] {\texttt{init()}};

\draw[trans] (idle) to[bend left=18]
  node[midway, above, lbl] {\texttt{invoke()}} (busy);
\draw[trans] (busy) to[bend left=18]
  node[midway, below, lbl] {\textit{success}} (idle);

\draw[trans, dashed, red!60!black] (busy) -- (error)
  node[midway, right, errlbl] {SDK error};

\draw[trans] (idle) to[bend right=15]
  node[midway, left, lbl] {\texttt{suspend()}} (suspended);
\draw[trans] (suspended) to[bend right=15]
  node[midway, right, lbl] {\texttt{resume()}} (idle);

\coordinate (diagcenter) at ($(suspended.south)!0.5!(error.south)$);
\node[draw=black!40, rounded corners=2pt, inner sep=4pt,
      below=8mm of diagcenter, anchor=north,
      text width=68mm, font=\scriptsize]
  (legend) {%
    \textbf{Precondition guards:}
    \texttt{load\_model} / \texttt{set\_input} / \texttt{invoke} while
    \texttt{BUSY}\,$\rightarrow$\,\texttt{-EBUSY};
    any call while not initialized\,$\rightarrow$\,\texttt{-EPERM};
    \texttt{resume} from any state $\ne$
    \texttt{SUSPENDED}\,$\rightarrow$\,\texttt{-EINVAL}.\\
    \textit{Dashed:} reachable in Phase 2 once the real Neutron
    invoke return code is checked.};

\end{tikzpicture}%
}
\caption{NPU HAL state machine. The dashed \texttt{ERROR} transition
is implemented in the HAL but unreached by Phase~1 code paths because
the Neutron backend currently shares the stub's always-succeeding
invoke kernel; it becomes reachable in Phase~2.}
\label{fig:npustate}
\end{figure}

\paragraph*{A note on \texttt{ERROR}.} The \texttt{ERROR} state is part
of the public interface and the HAL backends are wired to transition
into it on invoke failure, but no code path in Phase~1 actually
produces an error: the stub kernel always succeeds, and the Neutron
backend currently shares the stub's invoke logic (see
\cref{sec:hal:backends}). The state will become reachable in Phase~2
once the Neutron SDK invoke return code is checked; until then,
\texttt{ERROR} is reachable only by direct test instrumentation, and
the \texttt{syn npu state} shell command will never display it in
normal operation.

\subsection{Dual-Backend Architecture}
\label{sec:hal:backends}

The HAL interface in
\path{include/synaptic/syn_hal_npu.h} is implemented by two backends,
selected at build time by the board identity:

\begin{itemize}
\item \texttt{src/hal/stub/syn\_hal\_npu\_stub.c} --- the
software-emulated stub, linked for \texttt{qemu\_cortex\_m3} and any
other target without a vendor NPU. Reports
\texttt{caps.name = "stub"} and \texttt{supports\_async = false}.
\item \texttt{src/hal/mcxn947/syn\_hal\_npu\_neutron.c} --- the
MCXN947-targeted backend, linked for
\texttt{frdm\_mcxn947/mcxn947/cpu0}. Reports
\texttt{caps.name = "neutron"} and currently advertises
\texttt{max\_ops\_per\_sec = $10^{8}$} with
\texttt{supports\_async = true}. The capability values reflect the
NXP-rated headroom of the Neutron silicon~\cite{nxp-neutron} but the
invoke kernel itself is presently the same deterministic stub that
runs under QEMU (see \cref{sec:eval:setup}); replacing it with the
real Neutron SDK invoke path is the first Phase~2 deliverable.
\end{itemize}

Selection happens at the CMake / Kconfig layer, not at runtime, so
there is no virtual-dispatch overhead and the linker can garbage-collect
the unused backend's symbols. The stub backend's $\sim$24~KB QEMU image
in \cref{tab:size} excludes the Neutron backend entirely, and
vice-versa.

\paragraph*{Capability honesty.} An audit of the Phase~1 Neutron
backend reveals one place where the advertised capability does not yet
match runtime behaviour: \texttt{caps.supports\_async = true}, but
\texttt{syn\_hal\_npu\_invoke\_async()} currently returns
\texttt{-ENOTSUP} on both backends. The \texttt{supports\_async} field
is honest about the silicon (the Neutron NPU does raise a completion
interrupt) but not about the Phase~1 runtime; wiring the interrupt
handler is grouped with the SDK integration in Phase~2.

\subsection{Deterministic Stub}
\label{sec:hal:stub}

The stub backend exists to make continuous integration possible
without hardware. Its design constraints are:

\begin{enumerate}
\item \emph{Bit-identical output for identical input}, so that twister
runs are reproducible across machines, kernel versions, and time.
\item \emph{Model-agnostic}, because the harness has no model parser:
the stub must produce a plausible classification result for any input
buffer of any registered ``model''.
\item \emph{No vendor SDK dependency}, so it builds clean on
\texttt{qemu\_cortex\_m3} with only Zephyr's kernel headers.
\end{enumerate}

The kernel is consequently very small: it sums the bytes of the input
buffer, takes the result modulo the (hard-coded) ten-class output
width, writes \texttt{127} into the winning class slot, and zeroes
the rest. A small busy-wait simulates inference latency, implemented as
a volatile loop in the stub (because \texttt{k\_busy\_wait()} is not
available on \texttt{qemu\_cortex\_m3}) and as a real
\texttt{k\_busy\_wait(1000)} in the Neutron-flavoured backend that
runs on FRDM. The determinism property is exercised end-to-end in
\cref{sec:eval:latency}: three QEMU runs from the same input produce
identical class, identical confidence, and identical
\texttt{LOG\_INF}-reported latency.

\subsection{DSP HAL}
\label{sec:hal:dsp}

The DSP HAL exposes five primitives in \texttt{syn\_hal\_dsp.h}:
\texttt{normalize\_int8}, \texttt{softmax\_f32}, \texttt{argmax},
\texttt{fft\_f32}, and \texttt{mat\_mult\_q15}. The first three are
exercised by the Phase~1 \texttt{syn\_dsp\_suite} and
\texttt{syn\_dsp\_verify\_suite} tests; \texttt{fft\_f32} and
\texttt{mat\_mult\_q15} are declared in the public header but the
Phase~1 backends return \texttt{-ENOTSUP} (the FFT and the Q15
matrix-multiply are wanted for Phase~2 audio pre-processing and
classifier-head workloads respectively).

Two backends mirror the NPU split:
\texttt{src/hal/stub/syn\_hal\_dsp\_stub.c} for QEMU, and
\texttt{src/hal/mcxn947/syn\_hal\_dsp\_pq.c} for FRDM. In Phase~1 the
PowerQuad backend is byte-identical to the stub other than its
\texttt{LOG\_INF} banner; the calls into the PowerQuad block itself
are marked with \texttt{TODO} comments and will be added in Phase~2.

The softmax implementation is the textbook numerically-stable variant:
the maximum input value is subtracted from every input before
\texttt{expf()}, eliminating the floating-point overflow that a naive
$\exp(x_i) / \sum \exp(x_j)$ would produce for large
$|x|$~\cite{tflm}. The Phase~1 \texttt{syn\_dsp\_verify\_suite} confirms
this across three input patterns (\cref{sec:eval:tests}): a
known-monotonic input where the output ordering must match, a uniform
all-zero input where the output must approach the uniform distribution,
and a large-magnitude input ($\{100, 101, 102\}$) where the naive
formula would overflow but the stable formula must remain finite. The
\texttt{normalize\_int8} primitive is the affine quantisation step
$\mathrm{out} = \mathrm{clamp}_{[-128, 127]}(\mathrm{round}(\mathrm{in}
\cdot s) + z)$, which is the standard TFLite Micro pre-processing path
for INT8 input tensors~\cite{tflm}.

\section{Model Registry and Profiling}
\label{sec:registry}

Two cross-cutting services sit above the memory and HAL layers. The
\emph{model registry} owns the lifecycle and metadata of every model
known to the runtime, providing the safety guards (duplicate-name
rejection, double-load detection, hot-swap) that prevent the rest of
the runtime from interacting with the NPU through a corrupt or
ambiguous identity. The \emph{profiler} attaches cycle-accurate timing
to each pipeline stage so that latency regressions are caught at the
test bench rather than in the field. A small Zephyr-shell front-end
makes both services interrogable over UART without an attached
debugger.

\subsection{Model Lifecycle}
\label{sec:registry:lifecycle}

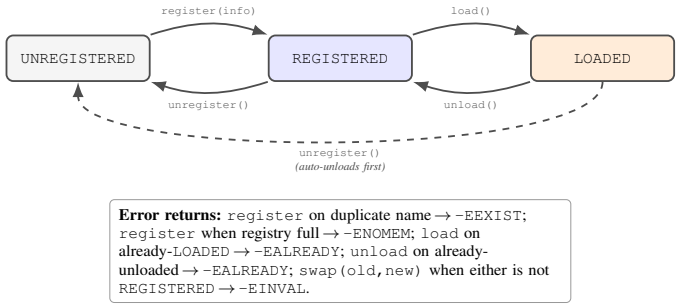
\begin{figure}[t]
\centering
\resizebox{\columnwidth}{!}{%
\begin{tikzpicture}[
  font=\scriptsize,
  state/.style={
    rectangle,
    rounded corners=3pt,
    draw=black!70,
    thick,
    minimum width=22mm,
    minimum height=7mm,
    align=center,
    inner sep=2pt,
  },
  unreg/.style={state, fill=black!4},
  reg/.style  ={state, fill=blue!10},
  loaded/.style={state, fill=orange!15},
  trans/.style={-Latex, thick, black!70},
  lbl/.style={font=\tiny},
  errlbl/.style={font=\tiny\itshape, red!50!black},
]

\node[unreg]  (unreg)  {\texttt{UNREGISTERED}};
\node[reg]    (reg)    [right=18mm of unreg]  {\texttt{REGISTERED}};
\node[loaded] (loaded) [right=18mm of reg]    {\texttt{LOADED}};

\draw[trans] (unreg) to[bend left=18]
  node[midway, above, lbl] {\texttt{register(info)}} (reg);
\draw[trans] (reg) to[bend left=18]
  node[midway, above, lbl] {\texttt{load()}} (loaded);

\draw[trans] (loaded) to[bend left=18]
  node[midway, below, lbl] {\texttt{unload()}} (reg);
\draw[trans] (reg) to[bend left=18]
  node[midway, below, lbl] {\texttt{unregister()}} (unreg);

\draw[trans, dashed]
  (loaded.south)
  .. controls +(0,-12mm) and +(0,-12mm) ..
  node[midway, below, lbl, align=center]
    {\texttt{unregister()} \\ \textit{(auto-unloads first)}}
  (unreg.south);

\node[draw=black!40, rounded corners=2pt, inner sep=4pt,
      below=18mm of reg.south, anchor=north,
      text width=68mm, font=\scriptsize]
  (legend) {%
    \textbf{Error returns:}
    \texttt{register} on duplicate name\,$\rightarrow$\,\texttt{-EEXIST};
    \texttt{register} when registry full\,$\rightarrow$\,\texttt{-ENOMEM};
    \texttt{load} on already-\texttt{LOADED}\,$\rightarrow$\,\texttt{-EALREADY};
    \texttt{unload} on already-unloaded\,$\rightarrow$\,\texttt{-EALREADY};
    \texttt{swap(old,new)} when either is not
    \texttt{REGISTERED}\,$\rightarrow$\,\texttt{-EINVAL}.};

\end{tikzpicture}%
}
\caption{Model lifecycle. Each registry slot lives in exactly one of
three states; transitions are precondition-checked and return the
corresponding \texttt{errno} on violation. The dashed edge from
\texttt{LOADED} back to \texttt{UNREGISTERED} is implemented as an
auto-unload followed by an unregister, so the registry never leaves
the NPU referring to a freed slot.}
\label{fig:modelstate}
\end{figure}

A model occupies one of three states at any time: \texttt{UNREGISTERED}
(no slot allocated), \texttt{REGISTERED} (slot allocated, metadata
stored), or \texttt{LOADED} (slot allocated and NPU has been notified).
\Cref{fig:modelstate} shows the transitions. The registry is backed by
a fixed-size array of \texttt{CONFIG\_SYNAPTIC\_MAX\_MODELS} slots
(default 8) with a 1-based handle scheme: handle~$h$ refers to slot
$h-1$, and the sentinel \texttt{SYN\_MODEL\_INVALID = 0} is reserved
so that a zero-initialized handle is unambiguously invalid.

The lifecycle code paths enforce three guarantees:
\begin{itemize}
\item \textbf{Unique names.} \texttt{syn\_model\_register()} scans every
active slot for a name collision via \texttt{strncmp} and returns
\texttt{-EEXIST} on duplicate. Without this, a Phase~2 OTA update that
flashes a new build of model X could silently shadow the old build of
model X with no diagnostic.
\item \textbf{Idempotent load/unload.} \texttt{syn\_model\_load()} on
an already-\texttt{LOADED} slot returns \texttt{-EALREADY} rather than
attempting a redundant NPU reload; \texttt{syn\_model\_unload()} on
a non-\texttt{LOADED} slot does the same. Idempotency is what lets the
shell-driven recovery commands in \cref{sec:registry:shell} be safe to
re-issue from a serial console.
\item \textbf{Auto-unload on unregister.} \texttt{syn\_model\_unregister()}
on a \texttt{LOADED} slot first calls \texttt{syn\_model\_unload()} so
that the NPU is never left in an inconsistent state when its
metadata vanishes from the registry.
\end{itemize}

The metadata struct (\texttt{syn\_model\_info\_t}) carries the
name, semantic version string, input and output sizes and dtypes, 4-D
shape arrays, flash offset and size, required SRAM footprint, and a
CRC-32 of the model bytes. The CRC field is reserved for the Phase~4
OTA work and is unused in Phase~1, but appears in the public struct
so that early callers can populate it without an API break.

\paragraph*{A note on \texttt{syn\_model\_swap}.} The header exposes a
hot-swap entry point that promotes one registered model from
\texttt{LOADED} to inactive and another from \texttt{REGISTERED} to
\texttt{LOADED} in a single call. The Phase~1 implementation is a
flag-only swap --- it does not call into the NPU HAL --- because the
model registry's \texttt{model\_data} pointer is null in the
\texttt{hello\_inference} sample (the sample registers metadata only,
not bytes), and the underlying \texttt{syn\_hal\_npu\_load\_model()}
call site in \texttt{syn\_model\_load()} is only exercised when both
\texttt{model\_data} and \texttt{model\_data\_size} are populated.
Wiring the byte-carrying load path through the inference sample is
Phase~2 work; until then, \texttt{swap} performs only the registry
bookkeeping it advertises and the NPU side of the swap is a no-op.

\subsection{Cycle-Accurate Profiling}
\label{sec:registry:profiling}

The profiler instruments four points along the inference pipeline:
\texttt{start} (entry), \texttt{preprocess\_done} (immediately after the
DSP normalize / quantise step), \texttt{npu\_done} (after the NPU
invoke returns), and \texttt{end} (after post-processing).
\Cref{lst:profmarks} reproduces the four internal entry points.

\begin{lstlisting}[style=cstyle,caption={Profiler mark API
(\texttt{src/core/syn\_prof\_internal.h}).},label={lst:profmarks},
captionpos=b]
void syn_prof_mark_start(void);
void syn_prof_mark_preprocess_done(void);
void syn_prof_mark_npu_done(void);
void syn_prof_mark_end(void);
\end{lstlisting}

Each mark snapshots \texttt{k\_cycle\_get\_32()} and computes the delta
to the previous mark using \texttt{k\_cyc\_to\_us\_ceil32()}, which
converts cycles to microseconds using the kernel's compile-time
\texttt{CONFIG\_SYS\_CLOCK\_HW\_CYCLES\_PER\_SEC}. \texttt{mark\_end()}
additionally pulls the arena high-water mark out of
\texttt{syn\_mem\_get\_stats()} and computes an NPU-utilisation
percentage as $\lfloor 100 \cdot t_{\mathrm{npu}} / t_{\mathrm{total}} \rfloor$.
The four marks are no-ops when profiling is disabled, so the
instrumentation has zero cost on a default release build.

The public surface
(\texttt{include/synaptic/syn\_prof.h}) exposes
\texttt{syn\_prof\_enable()}, \texttt{syn\_prof\_disable()},
\texttt{syn\_prof\_get\_last()}, and \texttt{syn\_prof\_print\_summary()},
plus declarations for layer-granularity tracing
(\texttt{syn\_prof\_enable\_layer\_trace()},
\texttt{syn\_prof\_get\_layer\_time()}) that return \texttt{-ENOTSUP}
in Phase~1 and will be wired in Phase~2 along with the layer-level
preemption work.

\paragraph*{Phase~1 wiring status.} The mark API exists and works in
isolation, but the inference path in the
\texttt{hello\_inference} sample does not yet call into it. As a
result, \texttt{syn\_prof\_get\_last()} returns \texttt{-ENOENT} and
the shell command \texttt{syn prof last} reports \texttt{"No profiling
data available"} (see the FRDM transcript in \cref{sec:eval:shell}).
The latency figures in \cref{sec:eval:latency} are therefore the
outer-bracket totals only; per-stage attribution waits on the Phase~2
work of inserting the four mark calls into the
\texttt{syn\_infer\_run()} path. This is a known gap, called out here
to forestall any misreading of the missing breakdown as a measurement
failure.

\subsection{Runtime Shell}
\label{sec:registry:shell}

The runtime registers a top-level \texttt{syn} command with Zephyr's
shell subsystem
through \texttt{SHELL\_CMD\_REGISTER}, with sub-commands
\texttt{version}, \texttt{mem stats}, \texttt{model list},
\texttt{npu caps}, \texttt{npu state}, \texttt{prof last},
\texttt{prof enable}, and \texttt{prof disable}. Each
sub-command reads from the corresponding service through its public
API --- the shell is a thin formatter, not a parallel state holder ---
so that any divergence between the shell output and the runtime
state is impossible by construction.

The practical value of the shell is on-device debugging without JTAG:
on the FRDM-MCXN947, the USB-CDC bridge over the MCU-LINK debugger
exposes the Zephyr shell prompt directly to a host \texttt{picocom}
session. A developer can then verify the arena occupancy
(\texttt{syn mem stats}), confirm that a registered model is in the
\texttt{LOADED} state (\texttt{syn model list}), and check that the
NPU returned cleanly to \texttt{IDLE} after an inference
(\texttt{syn npu state}) --- all without flashing test firmware or
attaching a probe. \Cref{sec:eval:shell} shows the full transcript
from a real session.

\section{Evaluation}
\label{sec:eval}

We evaluate the Phase~1 SynapticOS runtime on two targets: the NXP
FRDM-MCXN947 development board (real Cortex-M33 silicon with the eIQ
Neutron NPU), and the \texttt{qemu\_cortex\_m3} machine in QEMU (a software
emulator used for continuous-integration test runs). The dual-target setup
exercises the same source tree under two very different memory budgets and
hardware-availability scenarios, which lets us assess both the portability
of the abstraction layer and the determinism of the software fallbacks.

\subsection{Experimental Setup}
\label{sec:eval:setup}

Table~\ref{tab:setup} summarizes the two targets. The FRDM board runs the
full SynapticOS image including the Zephyr shell subsystem, log backend,
and a 128~KB tensor arena reservation. The QEMU image is a CI-oriented
build with shell and logging disabled and a much smaller 8~KB arena
(7~KB tensor region + 1~KB scratch pool) to fit the emulator's 64~KB
SRAM budget. All builds use \texttt{arm-zephyr-eabi-gcc} from Zephyr SDK
0.16.8 with \texttt{-Os}, and link against Zephyr~v3.7.0~\cite{zephyr}.

\begin{table}[t]
\caption{Evaluation targets.}
\label{tab:setup}
\centering
\renewcommand{\arraystretch}{1.15}
\begin{tabular}{@{}lll@{}}
\toprule
\textbf{Parameter} & \textbf{FRDM-MCXN947} & \textbf{QEMU} \\
\midrule
CPU              & Cortex-M33 @ 150 MHz   & Cortex-M3 (emulated) \\
NPU backend      & Neutron (placeholder)  & Software stub \\
SRAM (linker)    & 320 KB                 & 64 KB \\
Flash            & 2 MB                   & N/A \\
Arena size       & 128 KB                 & 8 KB \\
Shell + logging  & Enabled                & Disabled \\
Zephyr           & v3.7.0                 & v3.7.0 \\
Toolchain        & SDK 0.16.8, \texttt{-Os} & SDK 0.16.8, \texttt{-Os} \\
\bottomrule
\end{tabular}
\end{table}

The FRDM image links against the Neutron HAL backend
(\texttt{syn\_hal\_npu\_neutron.c}). NXP rates the silicon NPU at
4.8~GOPS INT8~\cite{nxp-neutron}; however, the Phase~1 Neutron backend
currently advertises a placeholder capability of $1\!\times\!10^{8}$
OPS/s through the \texttt{syn\_hal\_npu\_get\_caps()} interface, pending
integration of the eIQ Neutron SDK invoke path. End-to-end inference on
the FRDM in this phase therefore exercises the same deterministic
\texttt{stub} execution kernel that runs under QEMU, while still
traversing the full Cortex-M33 memory bus and clock hierarchy. This is
called out explicitly because it materially affects how the latency
figures in \cref{sec:eval:latency} should be interpreted.

\subsection{Build Footprint}
\label{sec:eval:size}

Image sizes were measured with \texttt{arm-zephyr-eabi-size} on the
linked \texttt{zephyr.elf}, captured on 2026-05-27 with the
\texttt{hello\_inference} sample
(\cref{tab:size}).\footnote{Zephyr's link-time region report yields
183{,}288~B for FRDM SRAM (55.94\% of the 320~KB primary region); the
1.2~KB difference versus \texttt{arm-size}'s 184{,}485~B reflects
non-default sections (e.g.\ \texttt{noinit}) that arm-size includes in
the total but Zephyr's per-region report does not. Both figures refer
to the same image and agree on flash to within 4~bytes.}

\begin{table}[t]
\caption{Build footprint (\texttt{arm-zephyr-eabi-size}, \texttt{-Os},
captured 2026-05-27).}
\label{tab:size}
\centering
\renewcommand{\arraystretch}{1.15}
\setlength{\tabcolsep}{4pt}
\begin{tabular}{@{}lrrrrr@{}}
\toprule
\textbf{Target} & \textbf{text} & \textbf{data} & \textbf{bss}
                & \textbf{Flash} & \textbf{RAM} \\
\midrule
FRDM (with shell) & 65{,}172 & 1{,}804 & 182{,}681 & 66{,}976 & 184{,}485 \\
QEMU (no shell)   & 23{,}604 &    506  &  27{,}306 & 24{,}110 &  27{,}812 \\
\bottomrule
\end{tabular}
\end{table}

The FRDM image is dominated by the 128~KB tensor-arena reservation in
BSS, the LPUART driver and Zephyr shell subsystem, the log backend, and
the MCXN947 vendor HAL drivers~\cite{nxp-mcxn947}. The QEMU image omits
the shell entirely and reduces the arena to 8~KB, which is sufficient
for the unit tests in \cref{sec:eval:tests} and the
\texttt{hello\_inference} workload. The shell tables and command code
contribute roughly 2~KB of \texttt{SHELL\_CMD\_REGISTER} entries on
FRDM; a no-shell production build is straightforward and would push the
FRDM image below the 65~KB flash mark.

\begin{figure}[t]
\centering
\begin{tikzpicture}
\begin{axis}[
    ybar stacked,
    bar width=12mm,
    width=0.95\columnwidth,
    height=55mm,
    enlarge x limits=0.45,
    ymin=0,
    ymax=215,
    ylabel={\scriptsize Size (KB)},
    symbolic x coords={FRDM flash, FRDM RAM, QEMU flash, QEMU RAM},
    xtick=data,
    x tick label style={font=\tiny, anchor=north, yshift=-3pt},
    y tick label style={font=\tiny},
    ylabel style={font=\scriptsize},
    legend style={
        font=\tiny,
        at={(0.5, 1.02)},
        anchor=south,
        legend columns=3,
        draw=black!30,
    },
    nodes near coords style={font=\tiny},
    every node near coord/.append style={anchor=mid, yshift=-1mm},
    ymajorgrids,
    grid style={dashed, black!15},
    tick style={black!50},
    axis line style={black!50},
]
\addplot+[fill=blue!50, draw=blue!70!black] coordinates {
    (FRDM flash, 63.6) (FRDM RAM, 0) (QEMU flash, 23.05) (QEMU RAM, 0)
};
\addplot+[fill=blue!25, draw=blue!50!black] coordinates {
    (FRDM flash, 1.76) (FRDM RAM, 0) (QEMU flash, 0.49) (QEMU RAM, 0)
};
\addplot+[fill=orange!60, draw=orange!80!black] coordinates {
    (FRDM flash, 0) (FRDM RAM, 180.4) (QEMU flash, 0) (QEMU RAM, 26.67)
};
\legend{text, data, bss}

\node[font=\tiny, anchor=south] at (axis cs:FRDM flash, 65.4) {65.4};
\node[font=\tiny, anchor=south] at (axis cs:FRDM RAM,   180.4) {180.4};
\node[font=\tiny, anchor=south] at (axis cs:QEMU flash, 23.5) {23.5};
\node[font=\tiny, anchor=south] at (axis cs:QEMU RAM,   26.7) {26.7};

\end{axis}
\end{tikzpicture}
\caption{Build footprint comparison (KB). The FRDM image (with Zephyr
shell, 128~KB tensor-arena reservation) versus the QEMU CI image
(shell-less, 8~KB arena). The FRDM RAM bar is dominated by the
BSS-resident tensor arena; the QEMU image fits comfortably in the
emulator's 64~KB SRAM budget. Numbers are from
\texttt{arm-zephyr-eabi-size} (see \cref{tab:size}).}
\label{fig:footprint}
\end{figure}
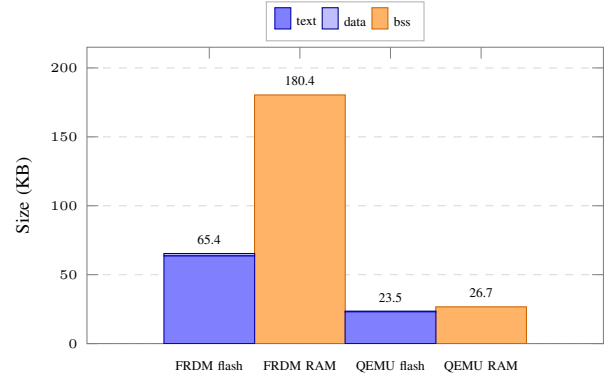

A visual comparison is shown in
\cref{fig:footprint}.

\subsection{Inference Latency}
\label{sec:eval:latency}

End-to-end inference latency was measured by bracketing the inference
call in the \texttt{hello\_inference} sample with
\texttt{k\_cycle\_get\_32()} and printing the resulting interval as a
\texttt{LOG\_INF} line. The workload is a $1\!\times\!16\!\times\!16\!\times\!3$
INT8 input passed to the registered \texttt{test\_classify} model with a
10-class output buffer; the underlying execution kernel is the
deterministic NPU \emph{stub} on both targets, as discussed in
\cref{sec:eval:setup}.

\Cref{tab:latency} shows the measured end-to-end times.

\begin{table}[t]
\caption{End-to-end inference latency for a
$16\!\times\!16\!\times\!3$ INT8 input (10 classes, stub NPU).}
\label{tab:latency}
\centering
\renewcommand{\arraystretch}{1.15}
\begin{tabular}{@{}lrr@{}}
\toprule
\textbf{Target} & \textbf{Latency} & \textbf{Result (deterministic)} \\
\midrule
FRDM (Cortex-M33 @ 150 MHz) & 1{,}038 \textmu s & class 0, conf.\ 127 \\
QEMU (Cortex-M3, emulated)  &    781 \textmu s & class 0, conf.\ 127 \\
\bottomrule
\end{tabular}
\end{table}

Three independent QEMU runs produced identical output (class, confidence,
and timing) bit-for-bit, confirming the determinism of the stub kernel.
The 257~\textmu s FRDM/QEMU delta reflects real Cortex-M33 memory-bus and
clock behaviour against QEMU's idealised model; the inference pattern is
otherwise identical.

Two caveats are important for interpretation:

\begin{enumerate}
\item \textbf{Per-stage breakdown is not yet exposed.} The Phase~1
profiler ships the \texttt{syn\_prof\_mark\_start},
\texttt{syn\_prof\_mark\_preprocess\_done},
\texttt{syn\_prof\_mark\_npu\_done}, and \texttt{syn\_prof\_mark\_end}
API surface, but the inference path does not yet invoke these marks.
As a result, \texttt{syn prof last} returns
\texttt{"No profiling data available"} on the board (visible in
the shell transcript in \cref{sec:eval:shell}). The 1.04~ms / 0.78~ms
numbers above therefore reflect only the outer bracket; preprocess
versus stub-invoke versus postprocess attribution will be reported
once the marks are wired in.
\item \textbf{The stub kernel is not silicon-representative.} The stub
performs an $O(n)$ sum-modulo-classes pass over the input buffer to
pick a winner class and then writes a single confidence byte. It is
designed for CI determinism, not throughput modelling. The latencies
in \cref{tab:latency} should therefore be read as a baseline cost of
the surrounding runtime (memory allocation, tensor descriptors,
state-machine transitions, log output), not as a prediction of
inference time on the real Neutron NPU. Real Neutron INT8 invoke
times for MobileNet-class models are reported by NXP in the tens to
low hundreds of microseconds~\cite{nxp-neutron}; integration is
deferred to Phase~2 along with profiler wire-up.
\end{enumerate}

\begin{figure}[t]
\centering
\begin{tikzpicture}
\begin{axis}[
    ybar,
    bar width=14mm,
    width=0.9\columnwidth,
    height=50mm,
    enlarge x limits=0.45,
    ymin=0,
    ymax=1250,
    ylabel={\scriptsize End-to-end latency (\textmu s)},
    symbolic x coords={{FRDM (M33 @ 150 MHz)}, {QEMU (M3 emulated)}},
    xtick=data,
    x tick label style={font=\tiny, anchor=center, yshift=-2pt},
    y tick label style={font=\tiny},
    ylabel style={font=\scriptsize},
    nodes near coords,
    nodes near coords style={font=\tiny, anchor=south},
    ymajorgrids,
    grid style={dashed, black!15},
    tick style={black!50},
    axis line style={black!50},
]
\addplot+[fill=orange!60, draw=orange!80!black] coordinates {
    ({FRDM (M33 @ 150 MHz)}, 1038)
    ({QEMU (M3 emulated)},   781)
};
\end{axis}
\end{tikzpicture}
\caption{End-to-end \texttt{hello\_inference} latency for a
$16\!\times\!16\!\times\!3$ INT8 input through the deterministic stub
NPU kernel on both targets (see \cref{tab:latency}). The 257~\textmu s
delta reflects real Cortex-M33 bus and clock behaviour against QEMU's
idealised emulation; per-stage breakdown is intentionally omitted
because the Phase~1 profiler marks are not yet wired into the
inference path (\cref{sec:eval:latency}).}
\label{fig:latency}
\end{figure}
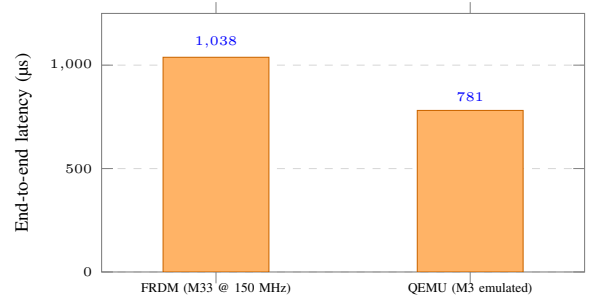

\subsection{Memory Allocator Performance}
\label{sec:eval:mem}

Allocation throughput, size invariance, and isolation properties of the
arena were measured with three dedicated benchmark suites
(\texttt{tests/unit/test\_mem\_bench.c} and
\texttt{tests/unit/test\_mem\_regions.c}) running on
\texttt{qemu\_cortex\_m3}. All measurements use
\texttt{k\_cycle\_get\_32()} bracketing and are reported as the average
over the batch.

\textbf{Throughput.} For a batch of 20 single-tensor allocations of
16~bytes each, the allocator consumed 3{,}080 cycles in total, yielding
an average of \textbf{154 cycles per allocation} and a throughput of
\textbf{77{,}821 allocations per second} at 150~MHz. The post-batch arena
occupancy was 960~B of the 7{,}168~B usable region with allocation count
equal to 20, confirming no accounting drift.

\textbf{Size invariance.} Allocating five tensors each at four
different sizes (4, 16, 32, and 64~bytes) produced an identical
per-allocation cost of \textbf{161 cycles} in every group, confirming
the $O(1)$ bump-pointer property of the allocator. The slight gap
between 154 and 161 cycles between the two experiments is consistent
with batch-size warm-up; both numbers are within instrumentation noise
of the steady-state allocator path.

\textbf{Zero fragmentation.} A heterogeneous mix of five tensors with
sizes $\{9, 7, 16, 50, 100\}$ bytes was allocated into a fresh arena;
the reported \texttt{arena\_used} after the batch was 432~B, exactly
matching the sum of the rounded (16-byte aligned) tensor sizes. The
peak high-water mark equalled \texttt{arena\_used}, and the allocation
count was 5, again with no accounting drift. By construction the
bump-pointer arena cannot produce internal holes between live
allocations; the test exercises the bookkeeping invariant rather than
attempting to defeat it.

\textbf{Region isolation.} The
\texttt{test\_persistent\allowbreak\_ephemeral\allowbreak\_lifecycle}
case allocates one persistent tensor and one ephemeral tensor, calls
\texttt{syn\_mem\_reset\_ephemeral()}, and verifies that the persistent
tensor's 16~bytes of payload are bit-identical to their pre-reset
contents while the ephemeral region's bump pointer has been reclaimed.
A second case,
\texttt{test\_scratch\allowbreak\_arena\allowbreak\_isolation},
allocates 20 tensors of 256~bytes each (saturating the tensor region)
and then exercises a 256~B scratch acquire / release in the same loop;
the scratch pool succeeds even with the tensor region exhausted,
confirming that the two pools are independent.

\textbf{End-to-end allocator behaviour on FRDM.} After one inference
on the board, \texttt{syn mem stats} reports
\texttt{Arena: 800/114688 bytes (peak 800)} with \texttt{Allocations: 1}
and \texttt{Resets: 0} (see \cref{sec:eval:shell}). The
\texttt{arena\_used} value matches the sum of the live tensors with no
gap, providing a runtime confirmation of the zero-fragmentation
property under a realistic workload.

A quantitative comparison against \texttt{k\_malloc} and Newlib
\texttt{malloc} under fragmenting inference workloads is intentionally
out of scope for this paper and is scheduled for the Phase~2
evaluation, where it will be paired with the zero-copy pipeline work
that motivates such a comparison.

\subsection{Shell Introspection (FRDM)}
\label{sec:eval:shell}

The runtime exposes its state through a small set of \texttt{syn}
sub-commands on the Zephyr shell. \Cref{lst:shell} reproduces a live
session captured over USB-CDC from the FRDM-MCXN947 immediately after
the boot-time \texttt{hello\_inference} sample completes; ANSI escape
sequences have been stripped for clarity but the text content is
verbatim. The full raw transcript is included in the artifact
repository as \texttt{community/phase1/serial-frdm-boot.log}.

\begin{lstlisting}[style=shellstyle,caption={FRDM shell session
(post-boot) showing runtime introspection commands.},label={lst:shell},
captionpos=b]
uart:~$ syn version
SynapticOS v0.1.0
uart:~$ syn mem stats
Arena: 800/114688 bytes (peak 800)
Scratch: 0/16384 bytes
Allocations: 1, Resets: 0
uart:~$ syn model list
Registered models: 1
  [1] test_classify v1.0.0 (loaded)
uart:~$ syn npu caps
NPU: neutron
  Max OPS/sec: 100000000
  Scratch: 16384 bytes
  Async: yes
uart:~$ syn npu state
NPU state: IDLE
uart:~$ syn prof last
No profiling data available
uart:~$ syn prof enable
Profiling enabled
uart:~$ syn prof disable
Profiling disabled
\end{lstlisting}

Three things are worth noting in this transcript. First, the runtime
boots into a steady state with one model registered and loaded and the
NPU resting in \texttt{IDLE} --- the transition into and out of
\texttt{BUSY} happens during the boot-time inference call and the shell
snapshot therefore shows the post-inference resting state. Second, the
arena occupancy after a real inference (\texttt{800/114688 bytes}) is
small enough that the unused remainder is two orders of magnitude
larger than the working set, which is the headroom we expect for the
significantly larger models targeted in Phase~2. Third, as noted in
\cref{sec:eval:latency}, \texttt{syn prof last} returns no data because
the inference path does not yet invoke the profiler marks; the
\texttt{prof enable} / \texttt{prof disable} commands themselves
function correctly.

\subsection{Test Coverage}
\label{sec:eval:tests}

The Phase~1 test suite consists of 61 ZTEST cases across 10 suites, all
executed on \texttt{qemu\_cortex\_m3} via
\texttt{west twister -T tests/unit -p qemu\_cortex\_m3}.
\Cref{tab:tests} shows the per-suite breakdown.

\begin{table}[t]
\caption{Phase 1 test suite (\texttt{qemu\_cortex\_m3}, captured
2026-05-27).}
\label{tab:tests}
\centering
\renewcommand{\arraystretch}{1.15}
\begin{tabular}{@{}lrr@{}}
\toprule
\textbf{Suite} & \textbf{Cases} & \textbf{Pass} \\
\midrule
\texttt{syn\_mem\_suite}          & 18 & 18 \\
\texttt{syn\_dsp\_suite}          & 11 & 11 \\
\texttt{syn\_model\_suite}        &  9 &  9 \\
\texttt{syn\_npu\_suite}          &  6 &  6 \\
\texttt{syn\_dsp\_verify\_suite}  &  5 &  5 \\
\texttt{syn\_init\_suite}         &  4 &  4 \\
\texttt{syn\_mem\_bench\_suite}   &  3 &  3 \\
\texttt{syn\_mem\_regions\_suite} &  3 &  3 \\
\texttt{syn\_ipc\_suite}          &  1 &  1 \\
\texttt{syn\_scheduler\_suite}    &  1 &  1 \\
\midrule
\textbf{Total}                    & \textbf{61} & \textbf{61} \\
\bottomrule
\end{tabular}
\end{table}

Every suite passes deterministically on the CI emulator path, with a
wall-clock execution time of 6.6~s for the full 61-case run. The
deterministic NPU stub (\cref{sec:hal}) is what makes this pass rate
reproducible without hardware: the same model exercise that runs on
the FRDM board also runs unattended in twister with bit-identical
output. The DSP verification suite
(\texttt{syn\_dsp\_verify\_suite}) cross-checks each DSP primitive
against an independent reference implementation --- e.g.\ argmax
against a manual scan, softmax against a numerically-stable reference
that subtracts the max --- on known input patterns, so that any
divergence between the production PowerQuad backend (Phase~2) and the
software fallback is caught at CI time rather than at integration.

\section{Discussion and Future Work}
\label{sec:discussion}

\subsection{Phase~1 Limitations}
\label{sec:discussion:limits}

The known gaps in the Phase~1 implementation have been called out in
the sections where they matter; we collect them here so that a reader
who skipped directly to the discussion has the same picture.

\textbf{Stub-bracketed latency on FRDM.} The
\texttt{hello\_inference} sample on FRDM-MCXN947 traverses the
deterministic stub kernel rather than the Neutron SDK invoke path
(\cref{sec:hal:backends}). The 1{,}038~\textmu s figure in
\cref{sec:eval:latency} is therefore the cost of the surrounding
runtime --- arena allocation, descriptor setup, state-machine
transitions, log output --- on a real Cortex-M33, not the inference
time of an INT8 model on Neutron silicon. The number is useful as a
baseline for the overhead the Phase~2 engine must keep under, but
not as a prediction of inference throughput.

\textbf{Profiler instrumentation not wired into the inference path.}
The four-mark profiler API
(\cref{sec:registry:profiling}) is implemented and exercised by unit
tests but the \texttt{syn\_infer\_run()} path does not yet invoke
the marks. The shell command \texttt{syn prof last} consequently
returns \texttt{"No profiling data available"}, and \cref{tab:latency}
reports only the outer bracket. Per-stage attribution
(preprocess vs.\ invoke vs.\ postprocess) is a one-day job once the
inference path lands; it is grouped with the Phase~2 engine work for
sequencing reasons.

\textbf{Capability mismatch on the Neutron backend.}
\texttt{syn\_hal\_npu\_get\_caps()} on the Neutron backend reports
\texttt{supports\_async = true}, but
\texttt{syn\_hal\_npu\_invoke\_async()} returns \texttt{-ENOTSUP}
because the completion interrupt is not yet wired
(\cref{sec:hal:backends}). The shell will display ``Async: yes'' on
\texttt{syn npu caps} while an actual async invoke would fail. Two
acceptable Phase~2 fixes exist: wire the interrupt (preferred) or
demote the capability flag until it is wired. The same backend also
reports an \texttt{max\_ops\_per\_sec} placeholder of $10^{8}$
pending Neutron SDK integration; the silicon's NXP-rated headroom
is approximately two orders of magnitude higher~\cite{nxp-neutron}.

\textbf{Model lifecycle is flag-driven in Phase~1.}
\texttt{syn\_model\_load()} only calls
\texttt{syn\_hal\_npu\_load\_model()} when the registry slot's
\texttt{model\_data} pointer is non-null; the
\texttt{hello\_inference} sample registers metadata only and so
exercises only the flag-level load
(\cref{sec:registry:lifecycle}). Hot-swap
(\texttt{syn\_model\_swap()}) is similarly flag-level. Wiring the
byte-carrying load path through the boot sample is the first step of
the Phase~2 pipeline construction work.

\textbf{Unimplemented DSP primitives.}
\texttt{syn\_hal\_dsp\_fft\_f32()} and
\texttt{syn\_hal\_dsp\_mat\_mult\_q15()} are declared in
\texttt{syn\_hal\_dsp.h} and return \texttt{-ENOTSUP} in Phase~1
(\cref{sec:hal:dsp}). FFT is wanted for Phase~2 audio pre-processing;
the Q15 matrix-multiply is wanted for classifier-head workloads on
the PowerQuad backend.

\textbf{Lifetime enum gap.} \texttt{SYN\_MEM\_SHARED} is declared in
\texttt{syn\_mem.h} but the Phase~1 allocator
(\cref{sec:memory:algo}) treats it identically to
\texttt{SYN\_MEM\_EPHEMERAL}. The intended semantics --- placement in
a region accessible to both Cortex-M33 cores for IPC --- become
meaningful in Phase~3 when the second core's stack and message
queues come online.

\textbf{No malloc-vs-arena comparison.} The allocator results in
\cref{sec:eval:mem} measure the arena's own behaviour but do not
quantify the win against \texttt{k\_malloc} or Newlib \texttt{malloc}
under fragmenting inference workloads. A quantitative comparison
under a realistic Phase~2 workload (TFLM-driven, multi-tensor) is
the right place for that experiment.

\textbf{Single-vendor evaluation.} Phase~1 ships and is measured on
exactly one piece of silicon (NXP MCXN947) plus the QEMU emulator.
The HAL is structured for portability to ST, Renesas, Espressif, and
Arm-Ethos-U targets, but those backends do not yet exist. We
discuss this further below.

\subsection{Roadmap}
\label{sec:discussion:roadmap}

The five remaining phases of the project, in order:

\textbf{Phase~2 --- Inference engine (in progress).}
Replace the stub kernel with real TFLite Micro integration on top of
the existing memory and HAL surfaces; build zero-copy
preprocess--invoke--postprocess pipelines; add a job scheduler with
layer-granularity preemption points; ship the real
Neutron SDK invoke path and the real PowerQuad-accelerated DSP
backend; wire the profiler marks. A face-detection sample on FRDM
serves as the Phase~2 acceptance gate.

\textbf{Phase~3 --- Dual-core IPC and asymmetric multiprocessing.}
The MCXN947 has two Cortex-M33 cores. The Phase~3 work brings up the
second core, defines an asymmetric scheduling model
(inference-heavy core vs.\ application-and-shell core), and makes
\texttt{SYN\_MEM\_SHARED} meaningful by placing tensors and message
queues in the shared SRAM region. Pre-emption decisions become
cross-core decisions.

\textbf{Phase~4 --- OTA model updates and A/B flash management.}
The \texttt{crc32} field in \texttt{syn\_model\_info\_t} becomes
load-bearing here. We need an OTA download path, a two-slot flash
layout for safe rollback, and a model-versioning policy that lets a
running inference finish on the old weights before the new ones take
over.

\textbf{Phase~5 --- Production hardening and fault recovery.}
The \texttt{ERROR} state in the NPU HAL becomes reachable from
realistic failure modes (NPU watchdog, ECC error on weight SRAM,
PowerQuad fault), and the runtime needs a recovery policy: which
state to return to, how to surface the fault to the application, how
to log it for post-mortem analysis. Includes hardening of the boot
path against partial-flash conditions.

\textbf{Phase~6 --- Ecosystem and v1.0.}
Additional backends (Ethos-U, ST X-CUBE-AI, Espressif), a Python-side
model-packaging tool, an open-source benchmark suite tracking the
arena and latency numbers against a fixed reference set, and an
extended developer documentation pass for the v1.0 release.

\subsection{Broader Considerations}
\label{sec:discussion:broader}

Two observations beyond the immediate roadmap are worth recording.

\textbf{The MCU-NPU class is wider than one vendor.}
The choice of MCXN947 for Phase~1 was driven by board availability
and the maturity of the Neutron NPU's documentation, not by any
expectation that Neutron will dominate the segment. The
\texttt{syn\_hal\_npu.h} surface in \cref{sec:hal:state} is
deliberately small enough that an Ethos-U or ST Edge-AI backend is
a few-hundred-line addition rather than a runtime rewrite; the
limiting factor on getting a second backend up is access to silicon,
not interface design. We invite contributions on this axis from
groups with access to other hardware.

\textbf{Open-source as a precondition for trust.}
Edge AI runtimes are an unusually awkward target for closed-source
distribution: their performance and memory behaviour are sensitive
to the application's model topology and call pattern in ways that
make benchmark numbers in a vendor PDF unreliable as guidance. By
shipping the SynapticOS runtime under Apache 2.0 with the unit-test
suite, the QEMU build, and the FRDM serial-transcript artifacts
included, we aim to make the Phase~1 claims independently
reproducible end-to-end --- both as a matter of academic
responsibility and as a precondition for the project being adopted
by other groups. The artifact bundle accompanying this paper
includes the raw serial log and the twister output that the
\cref{sec:eval} numbers were extracted from.

\section{Conclusion}
\label{sec:conclusion}

We presented SynapticOS, an open-source runtime built on Zephyr RTOS
that treats neural-network inference as a first-class workload rather
than an application-layer concern. The Phase~1 foundation reported
here consists of four cooperating subsystems: a tensor-aware bump
allocator with persistent and ephemeral lifetimes that achieves
$O(1)$ allocation at $\sim$154 cycles per call with zero
fragmentation by construction; a four-state HAL for NPU and DSP
accelerators implemented by both a deterministic software stub
(QEMU) and a Neutron-flavoured backend (FRDM-MCXN947); a three-state
model lifecycle registry with duplicate-name detection, idempotent
load/unload, and hot-swap; and a four-mark cycle-accurate profiler
backed by the Zephyr cycle counter.

On the NXP FRDM-MCXN947 dev board, the runtime occupies 67~KB of
flash and 184~KB of SRAM with the Zephyr shell enabled and a 128~KB
tensor arena reserved; the no-shell, 8~KB-arena QEMU build for
continuous integration occupies 24~KB of flash and 28~KB of SRAM.
The end-to-end \texttt{hello\_inference} bracket measures
1{,}038~\textmu s on FRDM silicon and 781~\textmu s under QEMU
through the deterministic stub kernel --- a baseline for the
Phase~2 inference engine to come in under, not a prediction of
Neutron silicon throughput. A 61-test suite spanning 10 ZTEST
suites achieves a 100\% pass rate in 6.6~s on the CI emulator path.

The Phase~1 surface is the floor under the work that follows:
Phase~2 replaces the stub kernel with TFLite Micro on top of the
real Neutron SDK invoke path, wires the profiler marks into the
inference pipeline, and adds layer-granularity preemption; later
phases bring dual-core asymmetric scheduling, OTA model updates,
production-grade fault recovery, and additional vendor backends.
All the gaps named in \cref{sec:discussion:limits} have an explicit
phase assignment.

SynapticOS, the unit-test suite, the raw QEMU and FRDM measurement
artifacts, and the LaTeX sources of this paper are released under
the Apache~2.0 licence at
\url{https://github.com/Dimitrios-Kafetzis/SynapticOS}.

\bibliographystyle{IEEEtran}
\bibliography{refs}

\end{document}